# Multi-time Scale Identification for Multi-energy Systems

Chao Yang[a]. Yucai Zhu[a,*]

[a] *College of Control Science and Engineering, Zhejiang University, Hangzhou, 310027 China*

**Abstract:** Multi-energy systems have been leaping forward for its various benefits, e.g., energy conservation and emission reduction. Coupling components are capable of transmitting energy from one time scale system to another time scale system, so the multi-energy system exhibits multi-time scale characteristic and broad bandwidth, thereby causing difficulties in dynamic modeling. In this work, two-time scale system identification is studied. A method is developed to solve the problem, which is uses signal pre-filtering and subtraction. The high and low frequency parts of the two-time scale system are identified separately and then combined to form the incorporated in parallel structure. The consistency of the method is proved and case studies are used to verify the effectiveness of the method.

## 1. Introduction

*1.1 Research background and related work*

In conventional energy management, different energy systems have been managed separately. In this situation, electricity grid, heating network, and natural gas network are managed by national grid, district heating companies and gas companies, resulting low overall efficiency of power systems. Over the past decade, as fueled by the higher demand for energy and increasingly serious environmental problems, different energy systems are urgently required to be integrated for energy utilities.

Multi-energy system has been developed for its wide benefits (e.g., energy conservation and emission reduction)[1]. Coupling components (e.g., combined heat and power units (CHP)[2], combined cooling, heating, and power units (CCHP)[3], as well as electric heating pumps) couple the systems together, turning the decentralized systems into an integrated system.

The response time varies significantly with system: the dynamic response of electrical power system is significantly fast (in seconds) since electric energy is transmitted at the speed of light; hydraulic process has a fast dynamic response (in minutes) as pressure is transmitted at the speed of sound; network thermal process exhibits a slow dynamic response (in minutes or hours) because thermal energy is transmitted at the speed of mass flow rate; the dynamic response of building temperature is significantly slow (in hours or days) for the considerable thermal capacity of buildings. Since the time scales of the above systems are significant different, if these systems are coupled, the integrated system is called a multi-time scale system.

As illustrated in Fig. 1.1 (A), coupling component transfers disturbance bidirectionally, therefore, the dynamic characteristics of the subsystems are affected not only by their own characteristics, but also by the coupling systems. As illustrated in Fig. 1.1 (B), to an identical disturbance, the responses of different systems are significant different as the time scale difference exist. Thus, coupling between different energy systems leads to complex multi-time scale characteristics of the integrated system[4], making it difficult to model multi-energy systems.

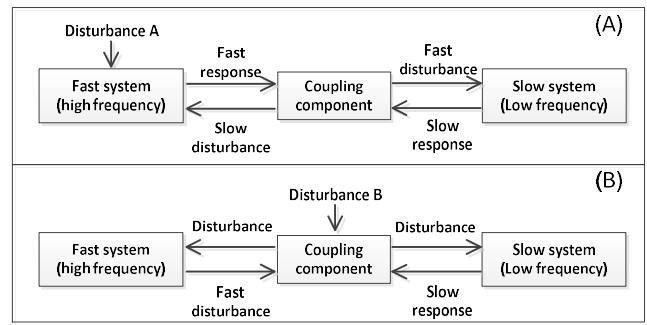

Fig. 1.1. Mechanism of multi-time scale systems

Dynamic modeling has been constantly neglected since the earlier studies primarily focused on planning and scheduling of multi-energy systems. Geidl[5] proposed a general steady-state modeling and optimization framework named 'Energy hub' for energy systems. Faeze[6] proposed a residential energy hub model to optimize the energy management of buildings. Bao[7] modeled the steady-state performance of some typical coupling components. The above-mentioned static models cannot express the multi-time scale characteristics of the system. Following many in-depth studies, researchers have realized the importance of dynamic modeling of multi-energy systems, and numerous relevant researches have been conducted.

---

* Corresponding author.
  E-mail: zhuyucai@zju.edu.cn (Y. Zhu), ychao@zju.edu.cn (C. Yang)

Mechanism models are capable of effectively explaining physical properties of multi-time scale systems, whereas the mechanism of a process is very complicated and costly to model. Thus, many scholars have proposed a range of simplified modeling methods. Whalley[8] proposed an equivalent dynamic modelling method for gas transportation via long pipelines. Chen[9] introduced an electrical analogy for analysis and optimization of heat network. Pan[4] studied the interactions of electricity and heating system based on a quasi-steady state model. Most of the above works concentrated on the modelling of network, the dynamic characters of the coupling components are overly simplified; Moreover, most of the results still need to be verified by practical applications.

The application of data-driven method has been increasingly emphasized in the field of multi-energy system modeling. Sheikhi[10] studied the demand side management of residential buildings based on reinforcement learning method. Fu[11] estimated the failure probability of gas supply based on data-driven model. The most common applications of data-driven methods in multi-energy systems are load forecasting and demand side management, whereas the method is rarely applied to dynamic modeling of multi-energy systems. Some literatures[12] highlighted that the main restrictions of data-driven modeling include the considerable amount of data required for modeling and the weak generalization ability of the model, which reveals the common problems of machine learning and other similar methods in industrial applications.

As another approach to process modeling, system identification has been extensively studied and widely applied in many fields[13,14]. Compared with mechanism modeling, system identification is in general cost effective and more accurate for complex dynamic systems. The system identification approach can reduce the amount of data required for modeling using proper identification test design. The generalization ability of the model can be enhanced by piecewise linearization of the model and other methods[15]. Besides, effective model validation methods exist to verify the obtained model.

However, there are few papers focusing on the identification of multi-energy systems. The multi-time scale character of multi-energy system may cause some difficulties when using conventional system identification methods.

*1.2 Contributions and outline*

This study focuses on dynamic modeling of multi-time scale systems using system identification, as the multi-time scale modeling is the critical problem in multi-energy system modeling[4]. The so-called filtering-subtraction method is developed to solve the problem, which is based on signal pre-filtering and subtraction. A simulated multi-energy system exhibiting multi-time scale character is used as a case study to verify the feasibility of the methods.

The rest of the paper is organized as follows. In Section 2, multi-time scale systems are introduced and the difficulties of identification are discussed and illustrated. In Section 3, the filtering-subtraction method is developed for multi-time scale system and verified by the simulation studies. In Section 4 the developed identification method is used to identify the dynamic model of a simulated industrial scale multi-energy system. Section 5 is the conclusion.

## 2. Issues in the identification of multi-time scale systems

The mentioned integrated system can be mathematically summarized as a multi-time scale system in accordance with automatic control theory. The multi-time scale system can be written as (2.1)[16]:

$$\begin{aligned} \dot{x} &= f(x,y,z) \\ \varepsilon \dot{y} &= g(x,y,z) \\ \mu \dot{z} &= h(x,y,z) \\ 1 &\gg \varepsilon \gg z \end{aligned} \quad (2.1)$$

where $x$, $y$ and $z$ are states vectors and $f$, $g$ and $h$ are functions of states $x$, $y$, and $z$.

This state space model indicates that the change rate of $x$ is significantly smaller than that of $y$ and $z$, exhibiting a noticeable multi-time scale character. As an example, the following continuous-time transfer function

$$G_A(s) = \frac{0.3}{0.019s^2 + 0.166s + 1} * \frac{15s+1}{30s+1} \quad (2.2)$$

where $s$ is the Laplace variable, shows a two-time scale behavior. The step response of $G_A(s)$ is illustrated in Fig. 2.1. The process has a fast response to the step input that reaches steady state in 1 second, and a slow response which reaches the steady state in about 100 seconds.

From this example, one can roughly define a two-time scale transfer function as a transfer function that contains two factors with very different speed or response time, typically with a difference of more than 30 times.

Such a multi-time scale behavior occurs in multi-energy systems. For instance, the response of back-pressure steam turbine power output to the change of input energy exhibits such multi-time scale characteristics.

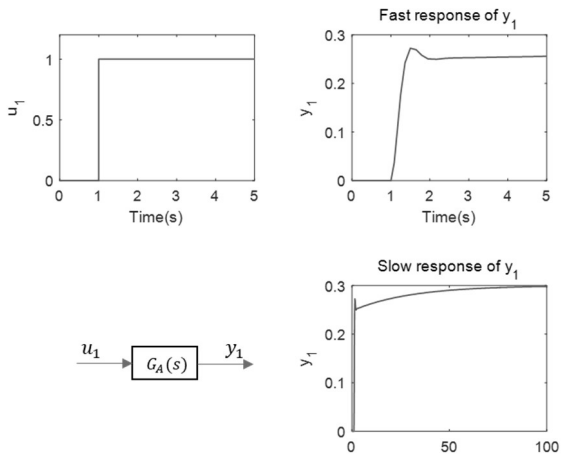

Fig. 2.1. Step response of $G_A(s)$

For a multi-input multi-output (MIMO) process, multi-time scale behavior can be shown using the following transfer function matrix.

$$G_B(s) = \begin{bmatrix} \dfrac{1}{(s+1)(100s+1)} & \dfrac{0.9}{2s+1} \\ \dfrac{1.2}{100s+1} & \dfrac{1.5}{200s+1} \end{bmatrix} \quad (2.3)$$

Here, entry (1, 1) is a two-time scale transfer function; entry (1, 2) is a fast transfer function; entries (2, 1) and (2, 2) are both slow transfer functions.

In system identification and model based control, it is more convenient to work with discrete-time models. Given a discrete-time single-input single-output (SISO) system as

$$y(t) = G(q)u(t) + v(t)$$
$$G(q) = \dfrac{B(q)}{A(q)} \quad (2.4)$$
$$v(t) = H(q)e(t) = \dfrac{C(q)}{D(q)}e(t)$$

where $u(t)$ is the input signal at sampling time $t$ and $y(t)$ is the output signal, $G(q)$ is the transfer operator, $\{v(t)\}$ is a stationary stochastic process with zero mean, $A(q)$, $B(q)$, $C(q)$ and $D(q)$ are polynomials of $q^{-1}$, $\{e(t)\}$ is a white noise sequence with zero mean and variance $\lambda$.

The model (2.4) is a specific parametrization of dynamic models which is called Box-Jenkins (BJ) model in identification literature; see Ljung[17] and Zhu[18].

Two methods of identification can be developed for the Box-Jenkins model in (2.4). The so-called *prediction error method* fully utilizes the model structure in (2.4) and calculates the model parameters of $A(q)$, $B(q)$, $C(q)$ and $D(q)$ by minimizing the following loss function

$$V_{PE} = \dfrac{1}{N}\sum_{t=n+1}^{N} \varepsilon_{PE}(t)^2 \quad (2.5)$$

where $N$ is the number of data and $n$ is the model order and
$$\varepsilon_{PE} = H^{-1}(q)[y(t) - G(q)u(t)] \quad (2.6)$$
is called the prediction error of model (2.4).

The second method is called output-error (OE) method that calculates the process model parameters of $A(q)$ and $B(q)$ by minimizing the following loss function

$$V_{OE} = \dfrac{1}{N}\sum_{t=n+1}^{N} \varepsilon_{OE}(t)^2 \quad (2.7)$$

where
$$\varepsilon_{OE} = y(t) - G(q)u(t) \quad (2.8)$$
is called the output error of model (2.4). Note that the output error method does not identify the disturbance model $H(q)$. It is easy to see that
$$\varepsilon_{PE} = H^{-1}(q)\varepsilon_{OE} \quad (2.9)$$

In general, the prediction error method is more accurate than the output error method. In technical terms, under open loop condition, a prediction error method is both consistent and efficient; an output error model is only consistent; see Ljung[17] and Zhu[18].

A two-time scale transfer function can be considered as two subsystems in series or parallel connection, one fast subsystem and one slow subsystem. The large difference in bandwidth can cause difficulties in system identification.

*Sampling time selection*

In accordance with the sampling theory, if the sampling time is too long, the information loss will occur. However, when the sampling time is too small compared with the system time constant, all poles are concentrated near point $z=1$ in the complex plane, thereby increasing the numerical sensitivity in identification calculation. Also fast sampling model may transform the system into a non-minimum phase system[19].

Normally, sampling time is in the range of 0.3% ~ 3% of the process settling time, a compromise choice for fast and slow sampling, which assure good numerical conditions, as well as generating sufficient dynamic information. For a multi-time scale system, selecting the sampling time is not an easy decision.

*Test signal design*

The test signal design refers to the signal spectrum design here. Since the spectrums of the disturbance signal and the control signal are determined by the system bandwidth, the signal frequency will show a direct association with the system bandwidth. It is difficult to design excitation signals that consider both the fast and slow subsystems for the wide frequency band of multi-time scale system.

**Example 2.1.** Given the system $G_A(s)$ in (2.2) which is a two-time scale system. Three identification tests are performed as follows.

(a) Fast excitation, fast sampling

(b) Slow excitation, slow sampling

(c) Superposed signal excitation, fast sampling

Output error method and predictive error (Box-Jenkins) method[17] are used to identify the models. For each test situation, 50 Monte Carlo simulations are used to show the results of identification.

GBN (Generalized binary noise) signals[20] with low-pass character are used as test signals. The sampling time and excitation signal are set according to the response time of different subsystems, and the test conditions are listed in Table 2.1.

Table 2.1. Conditions for the tests

| Test | Sampling time | GBN switching probability | Noise-to-signal ratio | Simulation Time |
|---|---|---|---|---|
| (a) | 0.04s | 0.06 | 15% | 4000s |
| (b) | 4 s | 0.06 | 15% | 4000s |
| (c) | 0.04s | Superposed signal of (a) and (b) | 15% | 4000s |

In the simulations, an unmeasured disturbance $v(t)$ is added at the output which is a filtered white noise as in (2.10), and its sampling time is that of the fast subsystem.

$$v(t) = \frac{\alpha(1-0.62q^{-1})}{1-0.92q^{-1}}e(t) \qquad (2.10)$$

where $e(t)$ is a white noise sequence and $\alpha$ is a constant adjusted so that the noise variance at the output is 15% of the measured output.

The true order is selected in the three experiments. The identified results are illustrated in Fig. 2.2, Fig. 2.3 and Fig. 2.4. Models obtained from all three test situations show some defects.

In situation (a), as shown in Fig. 2.2, the identified models are well geared into the fast part of the system, but the errors for the slow part are very large, which is due to the lack of low frequency excitation.

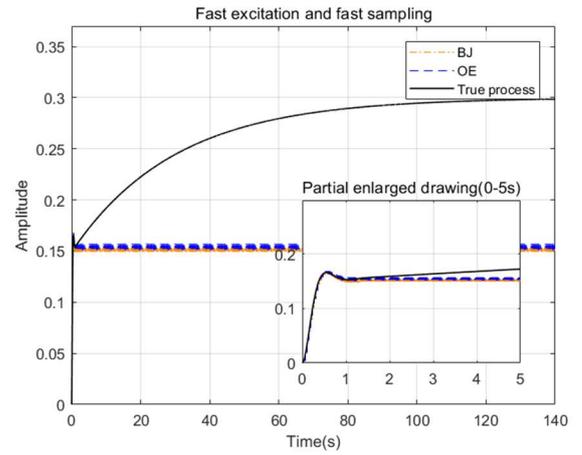

Fig. 2.2. Step response comparison of test (a).

In situation (b), as shown in Fig. 2.3, the identified models fit the slow part very well, but the fits for the fast part are very poor. This is caused by poor excitation at high frequencies.

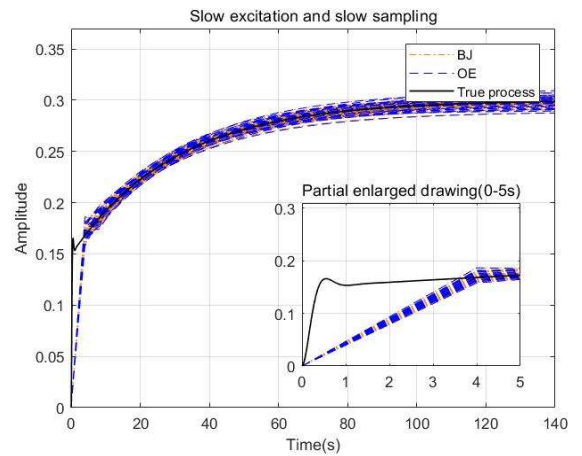

Fig. 2.3. Step response comparison of test (b).

In situation (c), a test signal that contains both fast excitation and slow excitation is used with fast sampling. One might expect that, with a proper test design, a good model could be identified. The identification results are plotted in Fig. 2.4 which is quite disappointing. Both the output error models and the predictive error models can fit the fast subsystem reasonably well, but the fits to the slow subsystem are rather poor.

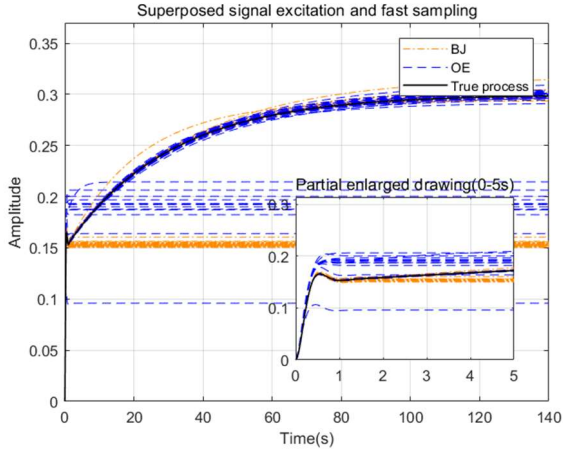

Fig. 2.4. Step response comparison of test (c).

The poor fit of the slow subsystem in situation (c) can be explained as follows. Because the two identification methods both minimize a single loss function, (2.5) or (2.7), the model errors of the fast subsystem and slow subsystem will interfere with each other. This is the case for situation (c): although both fast and slow test signals are used, the loss functions contain much more information of the fast subsystem than that of the slow subsystem, which makes the loss functions insensitive to the errors of the slow subsystem. In fact, as will be shown in Section 3, the fast subsystem model is also not optimal, due to the interference of the slow subsystem.

Based on the discussion, a way to reduce model errors in multi-time scale system identification is to reduce the interference between the fast and slow subsystems. In the sequel, methods of two-time scale system identification will be developed.

### 3. Two-time scale system identification

Given a continuous-time transfer function

$$G(s) = \frac{b_m s^m + b_{m-1} s^{m-1} + \ldots + b_0}{a_n s^n + a_{n-1} s^{n-1} + \ldots + a_0} \tag{3.1}$$

where $m > n$ and $n$ is called the order of the system. In order to keep the discussion simple, assume that system only has real poles, and the conclusion is also valid when there are complex poles.

Based on the partial-fraction expansion method[21], the transfer function can be expressed as the sum of single pole subsystems:

$$G(s) = \frac{b_m s^m + b_{m-1} s^{m-1} + \ldots + b_0}{a_n s^n + a_{n-1} s^{n-1} + \ldots + a_0}$$
$$= \frac{K_1}{s - p_1} + \frac{K_2}{s - p_2} + \ldots + \frac{K_n}{s - p_n} \tag{3.2}$$

where $p_j (j = 1, 2, \ldots, n)$ are the poles of the transfer function, $K_j (j = 1, 2, \ldots, n)$ are the coefficients correspond to the poles. It is well known that subsystem with a small pole is a slow subsystem and with a large pole a fast subsystem.

Now form the two-time scale system as

$$G(s) = G_{fst}(s) + G_{slw}(s) \tag{3.3}$$

where $G_{fst}(s)$ is the fast subsystem which lumps all the terms with fast (large) poles and $G_{slw}(s)$ is the slow subsystem which lumps all the terms with slow (small) poles. It is assumed that the gains of the two subsystems are in the same order of magnitude.

Convert the continuous-time system to discrete-time at a sampling time suitable for the fast subsystem; add the unmeasured disturbance at the output. Then the discrete-time two-time scale system can be expressed as

$$\begin{aligned} y(t) &= G(q)u(t) + v(t) \\ &= G_{fst}(q)u(t) + G_{slw}(q)u(t) + v(t) \\ &= \frac{B_{fst}(q)}{A_{fst}(q)} u(t) + \frac{B_{slw}(q)}{A_{slw}(q)} u(t) + \frac{C(q)}{D(q)} e(t) \end{aligned} \tag{3.4}$$

where $G_{fst}(q)$ and $G_{slw}(q)$ are the transfer operators of the fast subsystem and the slow subsystem respectively, $A_{fst}(q)$, $B_{fst}(q)$, $A_{slw}(q)$, $B_{slw}(q)$, $C(q)$ and $D(q)$ are polynomials of $q^{-1}$, all with degree $n$, $\{e(t)\}$ is a white noise sequence with zero mean and variance $\lambda$. This model structure is the Box-Jenkins model for two-time scale systems.

As discussed before, due to single loss function minimization in identification algorithms, the fast and slow subsystems in (3.4) interfere with each other that often cause large errors in one of the model. The idea here is to remove the interference by estimating the fast and slow subsystems separately and then sum them together. This approach is to convert the original single loss function minimization to two loss function minimizations, one for each subsystem.

Before starting, some assumptions will be introduced. Denote $\omega_{HF}$ as the cut-off frequency of the fast subsystem and $\omega_{LF}$ as that of the slow subsystem. For a two-time scale system we have $\omega_{HF} \gg \omega_{LF}$.

**Assumption A1:** The system (3.4) is stable.

**Assumption A2:** The orders of the fast and slow subsystems are known which equal to $n$.

**Assumption A3:** The cut-off frequencies $\omega_{HF}$ and $\omega_{LF}$ of the fast and slow subsystems are known.

The developed method will be called *filtering-subtraction method*, short named *fil-sub method* because two data filtering operations and one subtraction are used in the method.

**Step 1. Test Design**

- Design a test signal $u_{HF}(t)$ with a wide power spectrum covering the bandwidth of the fast subsystem. Then, design a test signal $u_{LF}(t)$ with a narrow power spectrum covering the bandwidth of the slow subsystem. Use a sampling time suitable for the fast subsystem. Add the two signals to form the test signal

$$u(t) = u_{HF}(t) + u_{LF}(t) \tag{3.5}$$

Then perform the identification test. Denote the obtained dataset as

$$Z^N := \{u(1), y(1), u(2), y(2), ..., u(N), y(N)\} \tag{3.6}$$

where $N$ is the number of data samples. Note that this is similar to case (c) in Example 2.1.

**Step 2. Estimate the Fast Subsystem Model**

- Remove the information of the slow subsystem by high-pass filtering the input-output data. Denote the filter as $L_{LP}(q)$ with a sufficient high order, the cut-off frequency of the high-pass filter is chosen more than twice the cut-off frequency of the slow subsystem $\omega_{LF}$.

$$\begin{aligned} u_{hp}(t) &= L_{HP}(q)u(t) \\ y_{hp}(t) &= L_{HP}(q)y(t) \end{aligned} \tag{3.7}$$

Denote the filtered dataset as

$$Z_{hp} := \{u_{hp}(1), y_{hp}(1), ..., u_{hp}(N), y_{hp}(N)\} \tag{3.8}$$

Note that the high-pass filtering also removes the low frequency disturbance at the output.

- Estimate an output error model of the fast subsystem using the filtered dataset $Z_{hp}$ in (3.8) by minimizing the output error loss function

$$\begin{aligned} V_{OE}^{hp} &= \frac{1}{N} \sum_{t=n+1}^{N_1} \varepsilon_{OE}^{hp}(t)^2 \\ &= \frac{1}{N} \sum_{t=n+1}^{N_1} [y_{hp}(t) - G(q)u_{hp}(t)]^2 \end{aligned} \tag{3.9}$$

where $n$ is the order of the fast subsystem. Denote the obtained fast subsystem model as

$$\hat{G}_{fst}(q) = \frac{\hat{B}_{fst}(q)}{\hat{A}_{fst}(q)} \tag{3.10}$$

Alternatively, one can estimate a Box-Jenkins model by minimizing the corresponding predictive error loss function.

**Step 3. Estimate the Slow Subsystem Model**

- Remove the information of the fast subsystem from the data by subtract the simulated output of fast subsystem from the output data

$$y_{slw}(t) = y(t) - \hat{G}_{fst}(q)u(t) \tag{3.11}$$

Denote the dataset with the subtracted output as

$$Z_{slw} := \{u(1), y_{slw}(1), ..., u(N), y_{slw}(N)\} \tag{3.12}$$

- Reduce the high frequency disturbance by low-pass filtering the input and subtracted output data. Denote the filter as $L_{LP}(q)$, the cut-off frequency of the filter is chosen the same as that of the high-pass filter in (3.7).

$$\begin{aligned} u_{lp}(t) &= L_{LP}(q)u(t) \\ y_{lp}(t) &= L_{LP}(q)y_{slw}(t) \end{aligned} \tag{3.13}$$

Denote the filtered dataset as

$$Z_{lp} := \{u_{lp}(1), y_{lp}(1), ..., u_{lp}(N), y_{lp}(N)\} \tag{3.14}$$

- Estimate an output error model of the slow subsystem using the filtered dataset $Z_{lp}$ in (3.14) by minimizing the output error loss function

$$\begin{aligned} V_{OE}^{lp} &= \frac{1}{N} \sum_{t=n+1}^{N} \varepsilon_{OE}^{lp}(t)^2 \\ &= \frac{1}{N} \sum_{t=n+1}^{N} [y_{lp}(t) - G(q)u_{lp}(t)]^2 \end{aligned} \tag{3.15}$$

Denote the obtained slow subsystem model as

$$\hat{G}_{slw}(q) = \frac{\hat{B}_{slw}(q)}{\hat{A}_{slw}(q)} \tag{3.16}$$

One can also estimate a Box-Jenkins model by minimizing the corresponding predictive error loss function.

**Step 4. Form the Two-time Scale System Model**

$$\hat{G}(q) = \hat{G}_{fst}(q) + \hat{G}_{slw}(q) \tag{3.17}$$

**Discussion on the Filtering-subtraction Method**

- Test design in Step 1 is to ensure that the dataset contains sufficient information of both fast and slow subsystems, which is crucial for a successful system identification.
- Step 2 is the single loss function minimization for the fast subsystem without the interference of the slow subsystem, as its information is removed from the data by the high-pass filtering. Moreover, the high-pass filtering also removes the low frequency disturbance, which increases model accuracy.
- Step 3 is the single loss function minimization for the slow subsystem without the interference of the fast subsystem as its information is removed from the data by subtraction in (3.10). In addition, the signal-to-noise ratio is increased by the low-pass filtering, which increases model accuracy.
- For an open loop test, the output error method can be used; for a closed-loop test, a Box-Jenkins method should be used in order to ensure the consistency of the model.
- One can also perform two tests, one using a test signal with a wide power spectrum, the other using a test signal with a narrow power spectrum. Then, estimate the fast sub-model using the first dataset as in Step 2; estimate the slow sub-model using the second dataset as in Step 3.

For the theoretical analysis of the filtering-subtraction method, more assumptions are given here.

**Assumption A4:** The identification test is performed in an open loop operation.

**Assumption A5:** The high-pass filter $L_{HP}(q)$ removes the energy of slow subsystem completely from the output signal, namely

$$L_{HP}(q)G_{slw}(q) = 0 \qquad (3.18)$$

**Assumption A6:** The test signal $\{u(t)\}$ and unmeasured disturbance $\{v(t)\}$ are stationary stochastic processes with zero means.

**Assumption A7:** The minimizations of the output error loss functions (3.9) and (3.15) both converge to their global minima.

Now it is time to show that the filtering-subtraction method can indeed identify the two-time scale system with consistency.

***Theorem 3.1.*** *Given the two-time scale system in (3.4) with assumptions A1-A7 hold. Then, the output error model obtained by the filtering-subtraction method is consistent, namely*

$$\begin{aligned} \hat{G}_{fst}(q) &\to G_{fst}(q) \text{ as } N \to \infty \\ \hat{G}_{slw}(q) &\to G_{slw}(q) \text{ as } N \to \infty \end{aligned} \qquad (3.19)$$

**Proof.** In Step 2 of the method, the high-pass filtering in (3.7) on the output signal gives

$$\begin{aligned} y_{hp}(t) &= L_{HP}(q)y(t) \\ &= L_{HP}(q)[G_{HF}(q)u(t) + G_{LF}(q)u(t) + v(t)] \\ &= L_{HP}(q)[G_{HF}(q)u(t) + v(t)] \end{aligned} \qquad (3.20)$$

The third equality (3.20) is due to Assumption A5 on the high-pass filter.

The output error loss function (3.9) for estimating the fast subsystem model becomes

$$\begin{aligned} V_{OE}^{hp} &= \frac{1}{N} \sum_{t=n+1}^{N_1} [y_{hp}(t) - \hat{G}(q)u_{hp}(t)]^2 \\ &= \frac{1}{N} \sum_{t=n+1}^{N_1} L_{HP}^2(q)[\Delta G_{fst}(q)u(t) + v(t)]^2 \end{aligned} \qquad (3.21)$$

where

$$\Delta G_{fst}(q) = G_{fst}(q) - \hat{G}(q) \qquad (3.22)$$

is the model error of the fast subsystems.

When the test signal $\{u(t)\}$ and the unmeasured disturbance $\{v(t)\}$ are stationary stochastic processes with zero means, denote $E$ as expectation, we have as $N \to \infty$

$$\begin{aligned} V_{OE}^{hp} &\to EV_{OE}^{hp} = L_{HP}^2(q)E[\Delta G_{fst}(q)u(t) + v(t)]^2 \\ &= L_{HP}^2(q)[\Delta G_{fst}^2(q)Eu^2(t) + Ev^2(t)] \end{aligned} \qquad (3.23)$$

The second equality holds because input $\{u(t)\}$ and the disturbance $\{v(t)\}$ are not correlated due to open loop test.

Now, when the model order is correct and when the minimization of the loss function converges to its global minimum, we have

$$[\Delta G_{fst}(q)]^2 \to 0 \text{ as } N \to \infty \qquad (3.24)$$

which implies

$$\hat{G}(q) \to G_{fst}(q) \text{ as } N \to \infty \qquad (3.25)$$

This proves the first line of (3.19).

Now consider the subtraction (3.11) in Step 3 of the method.

$$\begin{aligned} y_{slw}(t) &= y(t) - \hat{G}_{fst}(q)u(t) \\ &= [G_{slw}(q) + G_{fst}(q) - \hat{G}_{fst}(q)]u(t) + v(t) \\ &= [G_{slw}(q) + \Delta G_{fst}(q)]u(t) + v(t) \end{aligned} \qquad (3.26)$$

The output error loss function (3.15) for estimating the slow subsystem model becomes

$$\begin{aligned} V_{OE}^{lp} &= \frac{1}{N} \sum_{t=n+1}^{N_1} [y_{lp}(t) - \hat{G}(q)u_{lp}(t)]^2 \\ &= \frac{1}{N} \sum_{t=n+1}^{N_1} L_{LP}^2(q)\{[\Delta G_{slw}(q) + \Delta G_{fst}(q)]u(t) + v(t)\}^2 \end{aligned} \qquad (3.27)$$

where

$$\Delta G_{slw}(q) = G_{slw}(q) - \hat{G}(q) \qquad (3.28)$$

is the model error of the slow subsystems.

Then, we have when $N \to \infty$

$$\begin{aligned}V_{OE}^{lp} &\to EV_{OE}^{lp} \\ &= L_{LP}^2(q)E\{[\Delta G_{slw}(q) + \Delta G_{fst}(q)]u(t) + v(t)\}^2 \\ &= L_{LP}^2(q)E[\Delta G_{slw}(q)u_{lp}(t) + v(t)]^2 \\ &= L_{LP}^2(q)[\Delta G_{slw}^2(q)Eu_{hp}^2(t) + Ev^2(t)]\end{aligned} \qquad (3.29)$$

The second equality is due to the consistency of the fast sub-model as shown in (3.25); the third equality is due to open loop test.

Finally, when the model order is correct and when the minimization of the loss function converges to its global minimum, we have

$$[\Delta G_{slow}(q)]^2 \to 0 \text{ as } N \to \infty \qquad (3.30)$$

which implies

$$\hat{G}(q) \to G_{slw}(q) \text{ as } N \to \infty \qquad (3.31)$$

End of proof.

The proof is just adding the filtering and subtraction in the normal consistency proof of the output error models as in, e.g., Ljung[18]. The same approach can be used to prove the consistency of the predictive error method, although it is more complex. Here we will state the result for the prediction error method and skip the proof.

***Theorem 3.2.*** *Given the two-time scale system in (3.4) with assumptions A1-A3, A5-A7 hold. Then, the Box-Jenkins model obtained by the filtering-subtraction method is consistent, namely*

$$\begin{aligned}\hat{G}_{fst}(q) &\to G_{fst}(q) \text{ as } N \to \infty \\ \hat{G}_{slw}(q) &\to G_{slw}(q) \text{ as } N \to \infty \\ \hat{H}(q) &\to H(q) \text{ as } N \to \infty\end{aligned} \qquad (3.32)$$

Note that the open loop test condition A4 is not necessary for prediction error methods. The consistency here is expressed in terms of transfer operators of the system and the disturbance model, which is sufficient for applications in prediction, control and optimization. The consistency can also be expressed in terms of model parameters and the proof for the latter is just a simple extension of the proof here.

***Simulation Test***

The proposed filtering-subtraction method is verified using simulation data of the system $G_A(s)$ in (2.2). The test conditions are the same as in Table 2.1 (c) of Example 2.1.

Box-Jenkins method is used in Step 2 and Step 3 of method.

In total 50 simulations are conducted and the model step responses are compared with the true process as shown in Fig. 3.1. Compare this figure with Fig. 2.4, the model accuracy is increased considerably where models of both fast and slow subsystems are with good quality.

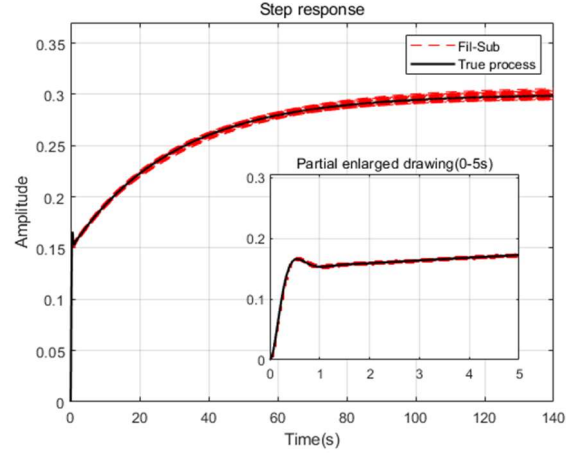

Fig. 3.1. True and model step responses

The accuracies of the identified models are also compared using noise-free data which is available in simulation study. The relative error (RE) is used to determine the model quality:

$$\text{RE} = \frac{\text{var}(y^o(t) - \hat{y}(t))}{\text{var } y^o(t)} \qquad (3.32)$$

where var(.) denotes the variance of the signal, $\hat{y}(t)$ is the simulated output, and $y^o(t)$ is the noise-free output. A smaller RE indicates a higher model accuracy.

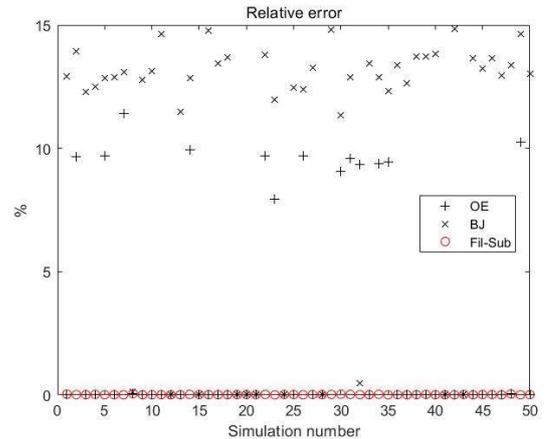

Fig. 3.2. REs of the OE method, the BJ method and the fil-sub method

Fig. 3.2 compares the REs of the three methods, which verifies that the filtering-subtraction method is superior to the normal output error (OE) method and Box-Jenkins (BJ) method.

# 4. Multi-energy system case study

## 4.1 System description

To verify the application of the algorithm in multi-energy systems, a CHP-based multi-energy system is implemented. The dynamic process simulation software Apros 6 is used for the study of the multi-energy system. Apros is a mature software product for dynamic simulation of power plants, energy systems and industrial processes, and it is widely used in related research works[22,23].

The major components of the system consist of a gas turbine (GT), a heat recovery boiler (HRSG), a backpressure steam turbine (ST), two district heaters, an auxiliary cooling tower (ACT) and two electric generators. The parameters of the main devices are shown in Table 4.1. The nominal power of the system is 40MW; the nominal heating area of the system is 1 million square meters.

Table 4.1. Parameters of main devices

| Device | Parameters |
|---|---|
| GT | Nominal power:30MW, nominal exhaust temperature:540 ºC |
| ST | Type: B10-3.0/0.07, medium temperature and medium pressure back-pressure steam turbine. Nominal power:10MW; nominal main steam parameter: 3.0MPa, 410 ºC; nominal exhaust steam pressure:0.07Mpa. |
| HRSG | Type: Q1000/540-72-3.15/415. Nominal capacity: 72t/h. |
| ACT | Rated inlet and outlet water temperature: 20/30 ºC. Maximum cooling power: 9MW |

Gas-steam combined cycle is used to supply energy: after the natural gas is burned, the flue gas drives the gas turbine to generate electricity; subsequently, it enters the heat recovery boiler to heat the feedwater, the generated steam drives the steam turbine to generate electricity. The back-pressure heating technology is adopted, heat power is supplied both by the exhaust steam of steam turbine (in district heater A) and the exhaust flue gas of gas turbine (in district heater B). The auxiliary cooling tower is used to assist in cooling the return water to prevent excessive back pressure of the steam turbine.

For this multi-energy system, three input variables ($m_{gas}, m_{air}, m_{BPF}$) and three input variables ($E_{ST}, E_{GT}, E_{BPF}$) are primarily discussed, as listed in Table 4.2. The three output variables are the major secondary energy outputs of the system, and the three input variables are the major control variables to control the secondary energy distribution of the system.

Table 4.2. Description of variables

| Tag | Description |
|---|---|
| $m_{gas}$ | Gas consumption rate, determining the total energy input to the system |
| $m_{air}$ | Air consumption rate, regulating the energy ratio between GT and ST by controlling the air-fuel ratio |
| $m_{BPF}$ | Bypass flow of district circulating water, regulating the backpressure of ST by controlling return water temperature of district heating system |
| $E_{ST}$ | Electric power of the steam turbine |
| $E_{GT}$ | Electric power of the gas turbine |
| $Q_H$ | Heat power supplied by the whole system |

The system is expressed as the following three-input three-out system

$$\begin{bmatrix} E_{ST} \\ E_{GT} \\ Q_H \end{bmatrix} = \begin{bmatrix} G_{11}(q) & G_{12}(q) & G_{13}(q) \\ G_{21}(q) & G_{22}(q) & G_{23}(q) \\ G_{31}(q) & G_{32}(q) & G_{33}(q) \end{bmatrix} \begin{bmatrix} m_{gas} \\ m_{air} \\ m_{BPF} \end{bmatrix} \quad (4.1)$$

where $G_{ij}(q)(i=1,2,3; j=1,2,3)$ is the transfer operator in discrete-time.

What needs to be emphasized is that the models in the matrix are all dynamic models, which contain more information compared to the conventional steady state models, e.g., energy hub models. The transfer operator matrix (4.1) not only contains the static energy conversion relationship (in gain information) which is critical to scheduling of the system, but also reflects the dynamic response characters of the system which is critical to control and optimization of the system.

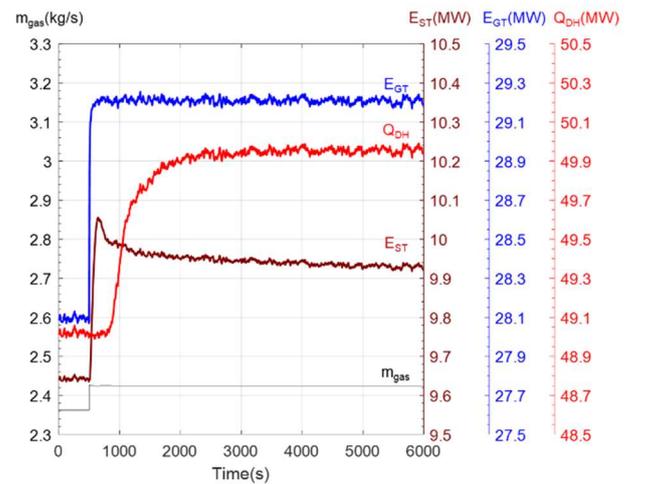

Fig. 4.2. Responses of $E_{ST}$, $E_{GT}$ and $Q_H$ to step change of $m_{gas}$

The multi-time scale character of the system can be illustrated by the responses of $E_{ST}$, $E_{GT}$ and $Q_H$ to a step change of $m_{gas}$ in Fig. 4.2.

Obviously, this system represents a typical multi-time scale system: The output signal $E_{ST}$ has both fast and slow responses to the input signal $m_{gas}$; the response speeds of output signals $E_{GT}$ and $Q_H$ to the input signal $m_{gas}$ are very different.

The mechanism of the multi-time scale character of the system illustrated in Fig. 4.3. For instance, if the load of the electric system changes, the automatic generation control (AGC) order is given to the unit, and the fuel quantity is adjusted according to the order, this is Stage 1. In stage 2, the output power of the gas turbine $E_{GT}$ changes fast, and the response time is about 70 seconds. In stage 3, as the thermal inertia of the steam turbine is relatively large, the output power of the steam turbine $E_{ST}$ and the supply water temperature change at medium speed, the response time is about 200 seconds. In stage 4, as the thermal inertia of the heating system is very large, the back-water temperature changes slowly; the backpressure of the steam turbine varies with the back-water temperature, and the electric power of steam turbine shows a direct association with the back pressure, then the power output of the steam turbine $E_{ST}$ changes very slowly, the response time is about 5000 seconds.

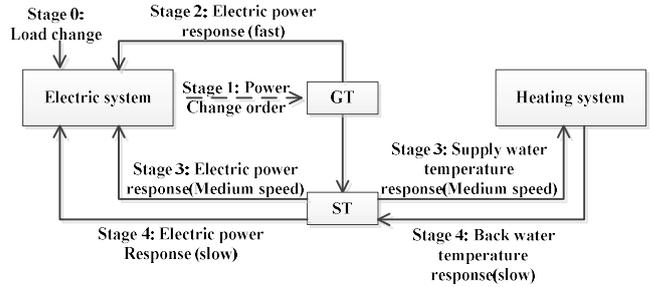

Fig.4.3. Mechanism of the multi-time scale character of the system

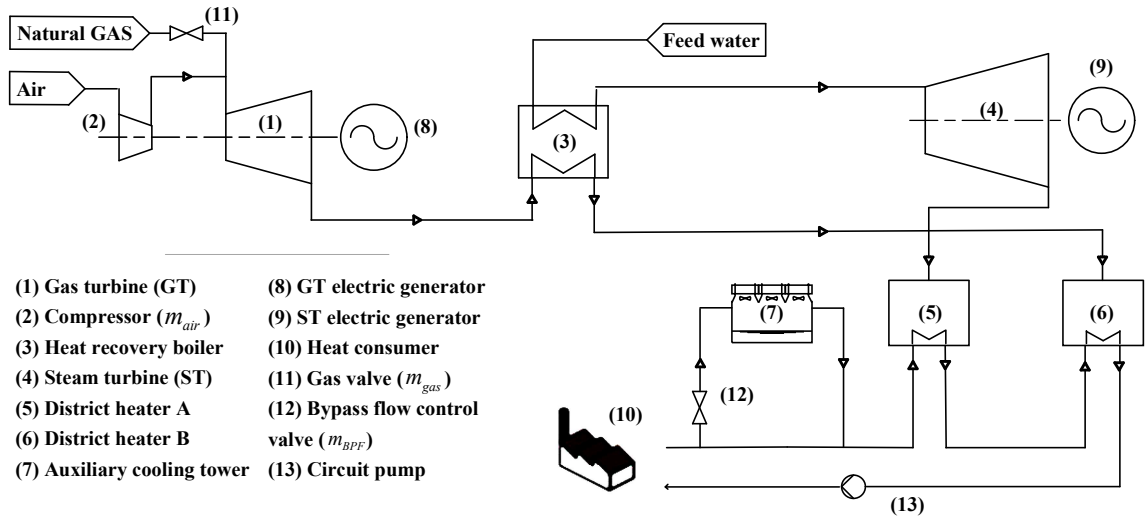

Fig. 4.1. Schematic diagram of the multi-energy system

## 4.2 Identification of dynamic model

The step test shown in Fig. 4.2 is performed before identification of dynamic model to basically estimate dominant process time constants, as listed in Table 4.3.

Table 4.3. Time constants of different channels

|  | $m_{gas}$ | $m_{air}$ | $m_{BPF}$ |
|---|---|---|---|
| $E_{ST}$ | 200 - 5000s | 200 - 5000s | 200 - 5000s |
| $E_{GT}$ | 70s | 70s | - |
| $Q_H$ | 2000s | 2000s | 2000s |

The test signals are designed based on the time constants of different channels listed in Table 4.4. Based on the time constants of different channels, the sampling time and the average switch time of GBN are determined.

Table 4.4. Conditions for the test of different channels

|  | Sampling time | Average switch time of GBN | Identification method |
|---|---|---|---|
| $E_{ST}$ | 10s | 100s/2000s Superposed signal | Fil-sub |
| $E_{GT}$ | 2s | 30s | BJ |
| $Q_H$ | 50s | 1000s | BJ |

Three tests are used for the identification of each output: (1) test with mixed fast and slow test signals for $E_{ST}$; see Fig. 4.4; (2) test with fast test signals for $E_{GT}$; see A part of Fig. 4.5; (3) test with slow test signals for $Q_H$; see B part of Fig. 4.5.

To make the simulations more realistic, the outputs of different channels are corrupted by filtered white noise as shown in (2.10), and the noise variance at the outputs is 5% of the measured outputs.

The proposed *fil-sub method* is used to identify the models relative to $E_{ST}$ ( $G_{11}(q), G_{12}(q), G_{13}(q)$ ) because there are two-time scale models in this channel. Box-Jenkins method is used in Step 2 and Step 3 of the *fil-sub method*. The output error (OE) method and predictive error (Box-Jenkins) method are also used here for comparison. In total 20 Monte Carlo simulations are used to show the results of identification.

The final output error (FOE) criterion[24] is used to determine the model orders which is defined as:

$$FOE = \frac{N+d}{N-d} \frac{1}{N} \sum_{t=1}^{N} [y(t) - \hat{G}(q)u(t)]^2 \qquad (4.2)$$

where $N$ is the number of estimation data samples, $d$ is the number of model parameters to be estimated, $y(t)$ is the measured output, $\hat{G}(q)$ is the identified model and $u(t)$ is the input. The model order with the minimal FOE used.

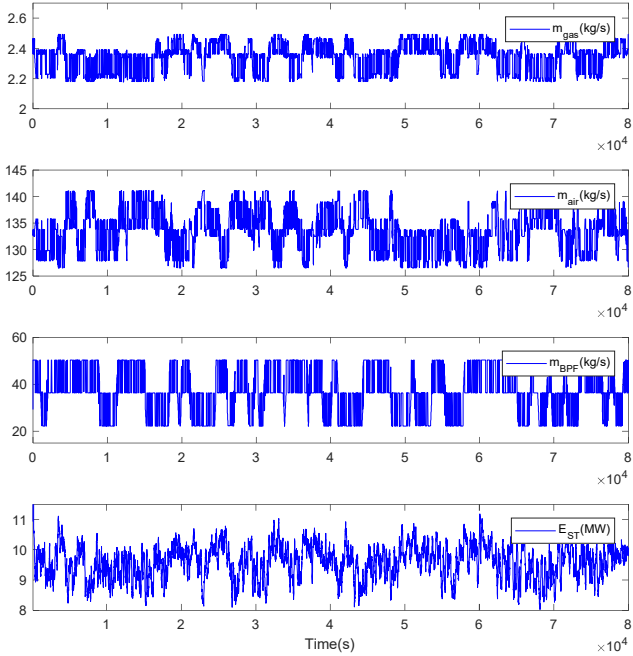

Fig. 4.4. Identification data for models related to $E_{ST}$

BJ method is used to identify models for $E_{GT}$ ( $G_{21}(q), G_{22}(q), G_{23}(q)$ ) and $Q_H$ ( $G_{31}(q), G_{32}(q), G_{33}(q)$ ) because these two channels contain only single-time scale models

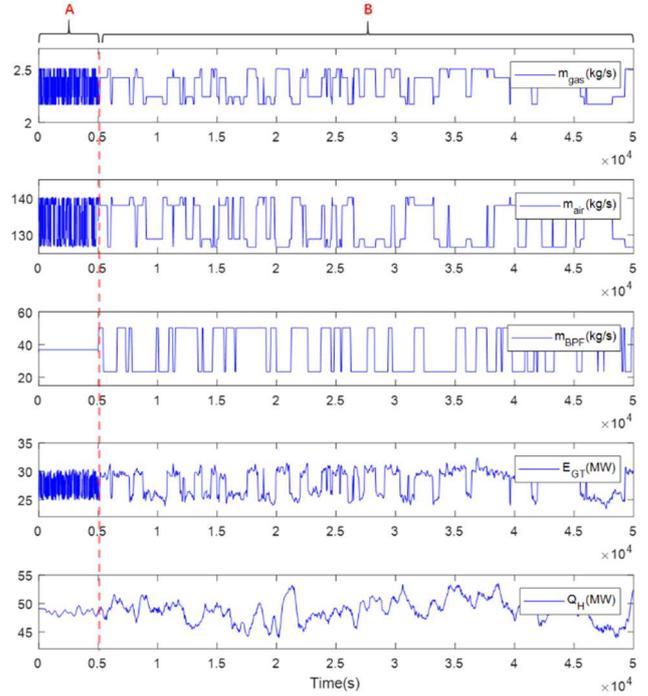

Fig. 4.5. Identification data for models related to $E_{GT}$ and $Q_H$: Data A (for models relative to $E_{GT}$) and Data B (for models relative to $Q_H$)

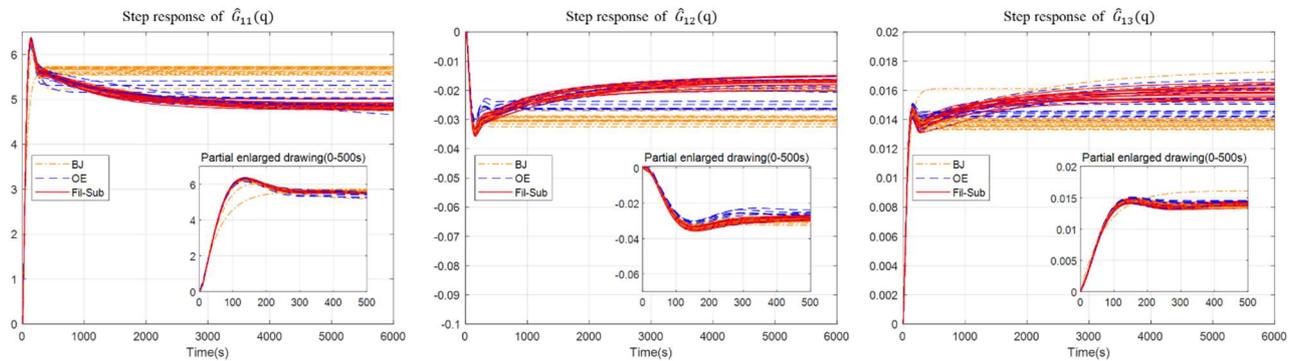

Fig. 4.6 Step response comparison of the identified dynamic model relative to $E_{ST}$ ( $\hat{G}_{11}(q)$ , $\hat{G}_{12}(q)$ , $\hat{G}_{13}(q)$ )

*Identification results*

Models of the three methods for $E_{ST}$ are compared since they have both fast and slow dynamics. The identification results of the models related to $E_{ST}$ are presented in Fig. 4.6. In 20 simulations, the output error (OE) models and the predictive error (BJ) models cannot fit the slow subsystem well; the models of the *fil-sub method* are much more accurate.

Step responses of the identified models related to $E_{GT}$ and $Q_H$ are presented in Fig. 4.7 which have either fast or slow dynamics.

Identified model fits of $E_{ST}$ (*fil-sub method*), $E_{GT}$ and $Q_H$ are presented in Fig. 4.8, Fig. 4.9 and Fig. 4.10, respectively. Validation data is used here to ensure no overfitting problem in the identified models.

The relative errors of $E_{ST}$, $E_{GT}$ and $Q_H$ are 5.66%, 5.90% and 5.75% respectively, which are close to the sizes of the disturbances. These good fits indicate good model quality.

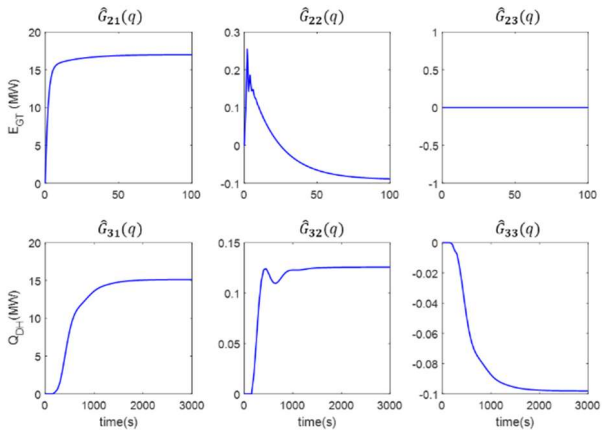

Fig. 4.7. Step response of the identified dynamic models relative to $E_{GT}$ ($G_{21}(q), G_{22}(q), G_{23}(q)$) and $Q_H$ ($G_{31}(q), G_{32}(q), G_{33}(q)$)

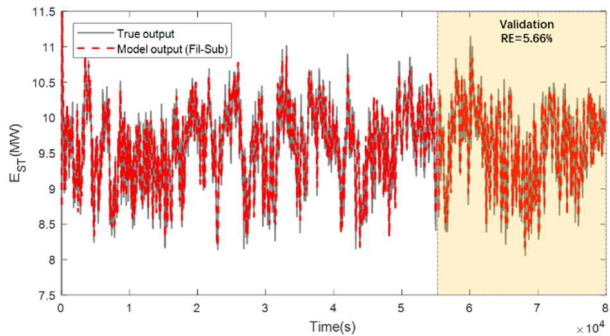

Fig. 4.8. $E_{ST}$ fitting of identified model

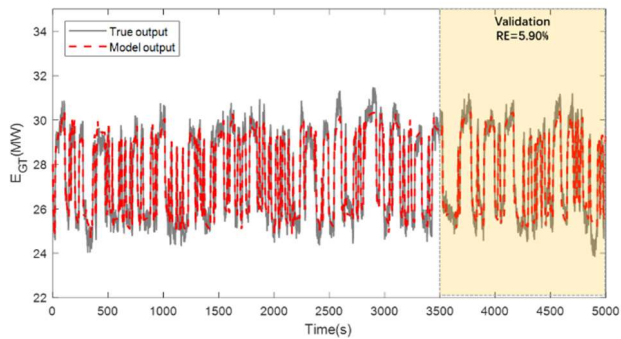

Fig. 4.9. $E_{GT}$ fitting of identified model

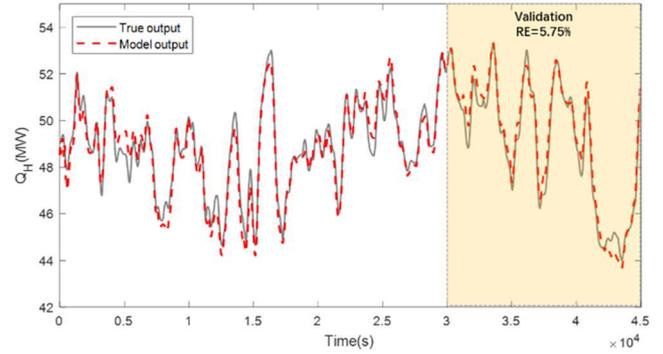

Fig. 4.10. $Q_H$ fitting of identified model

## 5. Conclusion

In this study, a multi-time scale system identification method is developed. The interference between the fast components and slow components are removed using proper test design, data pre-filtering and subtraction. The consistency of the method is established. The performance of the method is verified with simulation examples. For multi-energy systems, the availability of dynamic models is very promising as it paves the way towards advanced control, optimization and fault diagnosis. The proposed method is not limited to multi-energy systems and it can be used in other systems with multi-time scale characteristics.

### Acknowledgements

This work is supported by National Key R&D Program of China [grant number 2017YFA0700300] and Natural Science Foundation of China [grant number U1809207].